\documentstyle[11pt,newpasp,twoside,epsf]{article}
\def\edcomment#1{\iffalse\marginpar{\raggedright\sl#1\/}\else\relax\fi}
\reversemarginpar

\begin{document}

\title{Extrasolar Planets and Mean-Motion Resonances}
\author{Man Hoi Lee \& S.~J.~Peale}
\affil{Department of Physics, University of California, Santa Barbara,
       CA 93106, U.S.A.}

\begin{abstract}
The 2:1 orbital resonances of the GJ\,876 system can be easily established
by the differential planet migration due to planet-nebula interaction.
Significant eccentricity damping is required to produce the observed
orbital eccentricities.
The geometry of the GJ\,876 resonance configuration differs from that
of the Io-Europa pair, and this difference is due to the magnitudes of the
eccentricities involved.
We show that a large variation in the configuration of 2:1 and 3:1
resonances and, in particular, asymmetric librations can be expected among
future discoveries.
\end{abstract}

\section{Introduction}

Marcy et al. (2001) have discovered two planets in 2:1 orbital resonances
about the star GJ\,876.
This and possibly two other resonant pairs, one in 2:1 resonances about
HD\,82943 (Mayor et al. 2001; Go\'zdziewski \& Maciejewski 2001) and the
other in 3:1 resonances about 55\,Cnc (Marcy et al. 2002), and the ease of
capture into these resonances from nebula induced differential migration
of the orbits (e.g., Lee \& Peale 2002), mean that such resonances are
likely to be ubiquitous among extrasolar planetary systems.

\section{The GJ\,876 System}

A dynamical fit to the radial velocity data (e.g., Laughlin \& Chambers
2001) places the GJ\,876 planets on coplanar orbits deep in three
resonances at the 2:1 mean-motion commensurability.
The libration of the mean-motion resonance variables, $\theta_1 = \lambda_1
- 2\lambda_2 + \varpi_1$ and $\theta_2 = \lambda_1 - 2\lambda_2 + \varpi_2$,
and the secular resonance variable, $\theta_3 = \varpi_1 - \varpi_2 =
\theta_1 - \theta_2$, about $0^\circ$ (where $\lambda_{1,2}$ are the mean
longitudes of the inner and outer planet and $\varpi_{1,2}$ are the
longitudes of periapse) differs from the familiar geometry of the 2:1
resonances between Jupiter's satellites Io and Europa, where $\theta_1$
librates about $0^\circ$ but $\theta_2$ and $\theta_3$ librate about
$180^\circ$.
We have found that this difference is mainly due to the magnitudes
of the orbital eccentricities involved ($\langle e_1\rangle < 0.005$ for
the Io-Europa system, but $\langle e_1\rangle\approx 0.26$ for GJ\,876).
Stable simultaneous librations of the resonance variables require the
equality of the average periapse motions, $\langle \dot\varpi_1 \rangle =
\langle \dot\varpi_2 \rangle$.
When the $e_j$ are small, the $\dot\varpi_j$ are dominated by the lowest
order resonant terms in the disturbing potential, and $\langle \dot\varpi_1
\rangle = \langle \dot\varpi_2 \rangle$ requires $\theta_1 \approx 0^\circ$
and $\theta_2 \approx 180^\circ$.
On the other hand, when the $e_j$ are large, the dominance of
higher order terms means that $\langle \dot\varpi_1 \rangle = \langle
\dot\varpi_2 \rangle$ requires both $\theta_1$ and $\theta_2 \approx
0^\circ$ (see Lee \& Peale 2002 for details) and for some systems with
masses different from those in GJ\,876, $\theta_1$ and $\theta_2$ far from
either $0^\circ$ or $180^\circ$ (see \S~3).

\begin{figure}
\plottwo{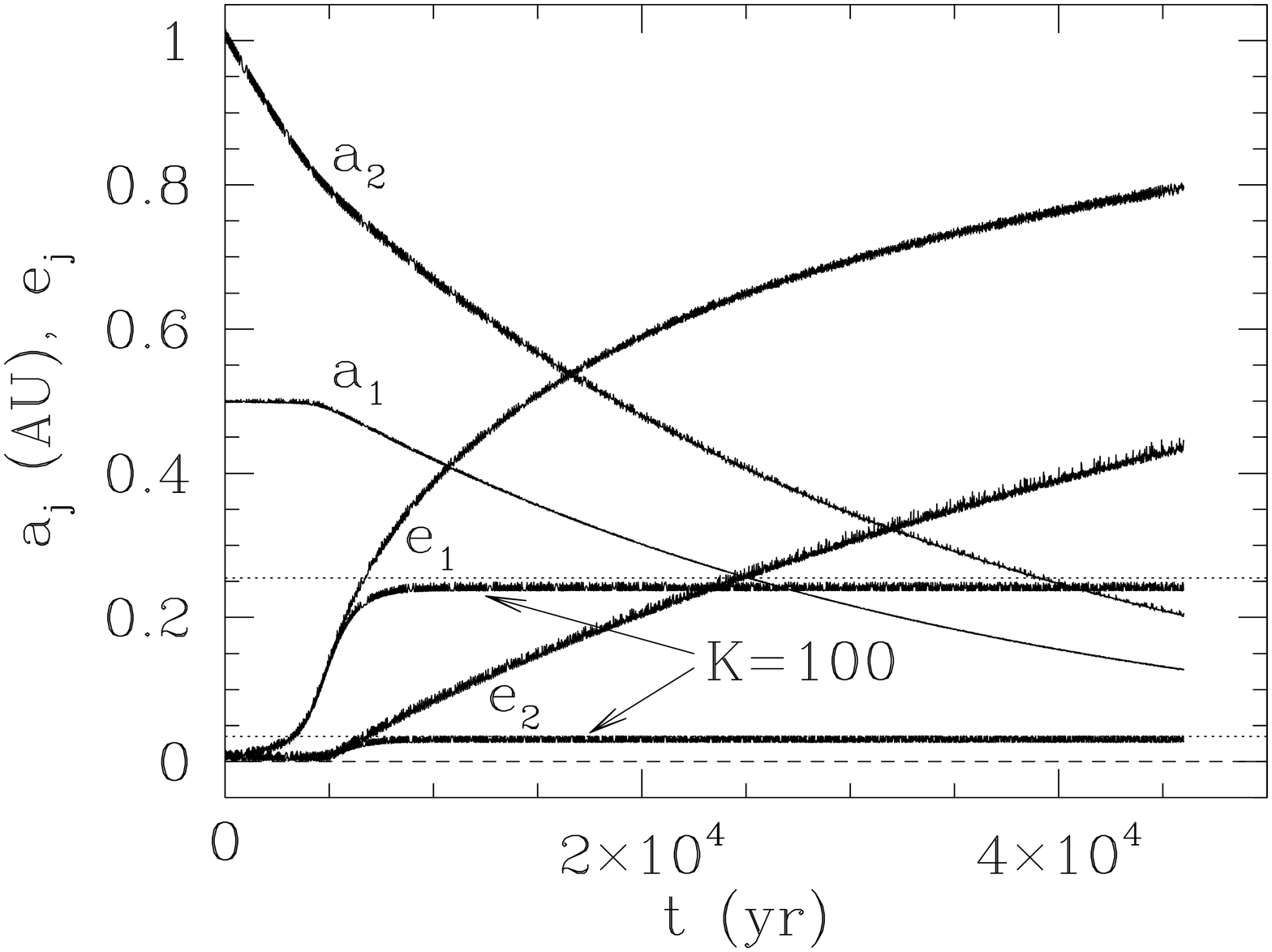}{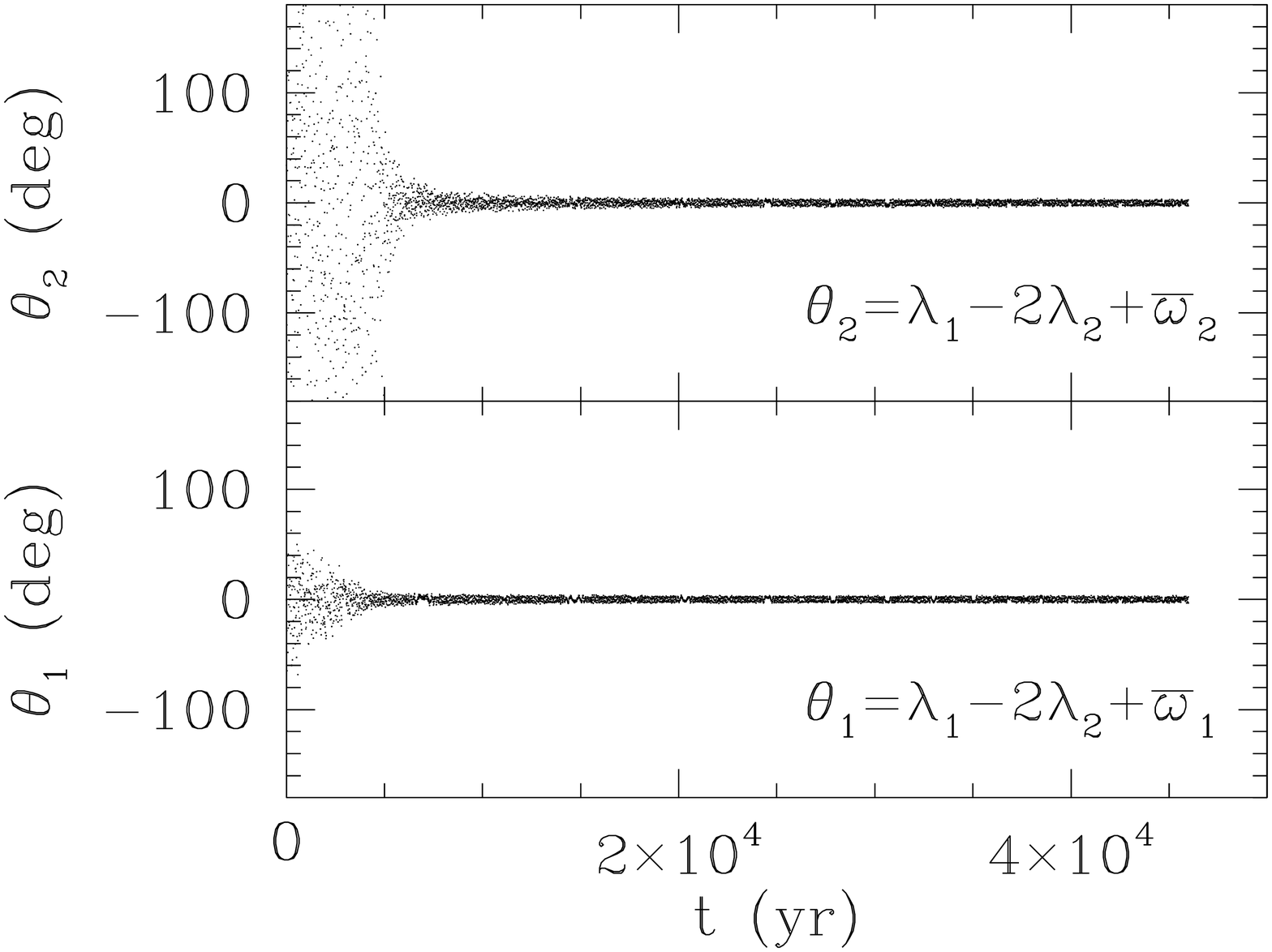}
\caption{Evolution of the GJ\,876 planets into 2:1 resonances for a
calculation where the outer planet is forced to migrate inward and
there is no eccentricity damping (see text for the curves labeled
$K = 100$).}
\end{figure}

\begin{figure}
\plottwo{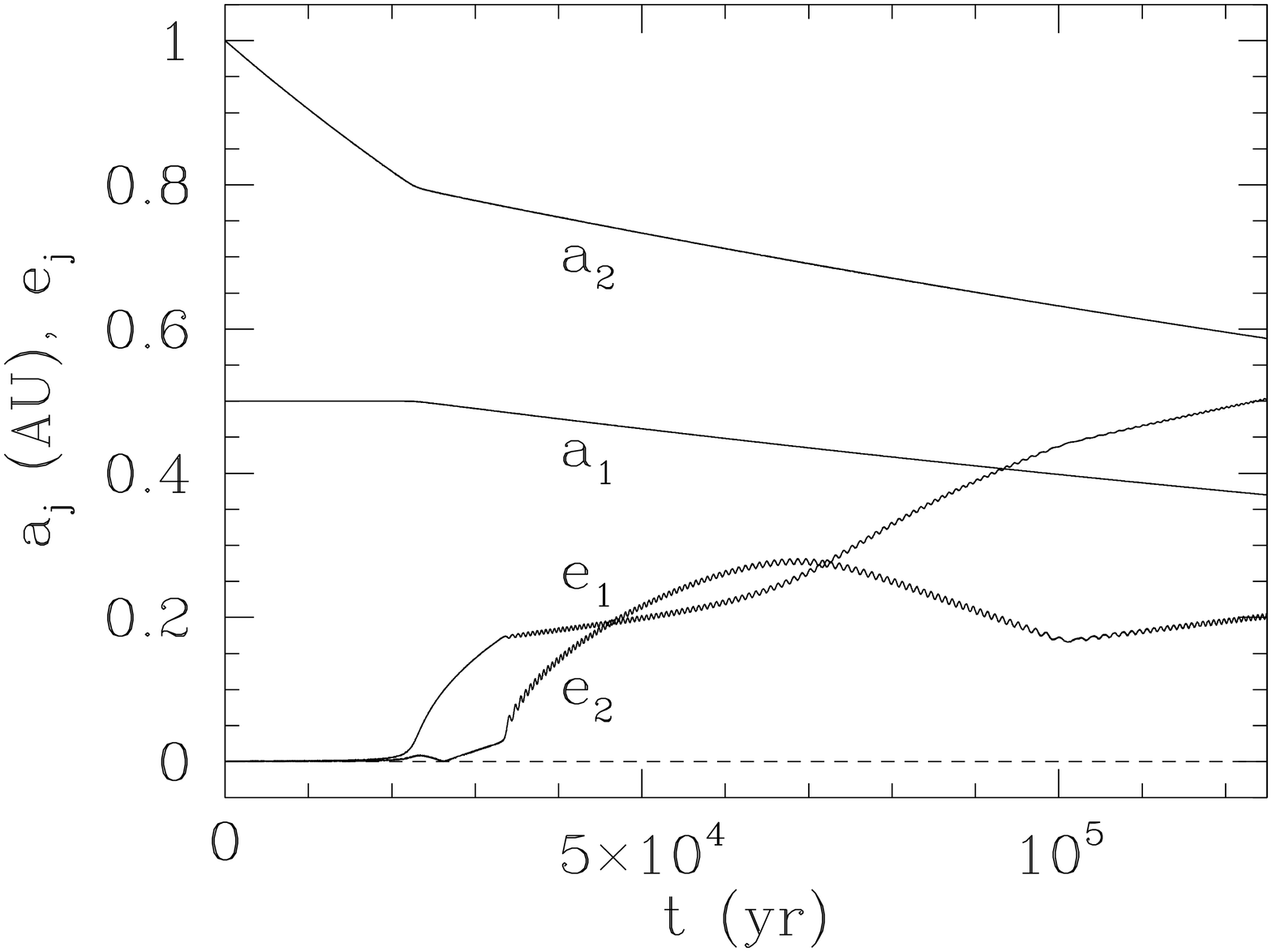}{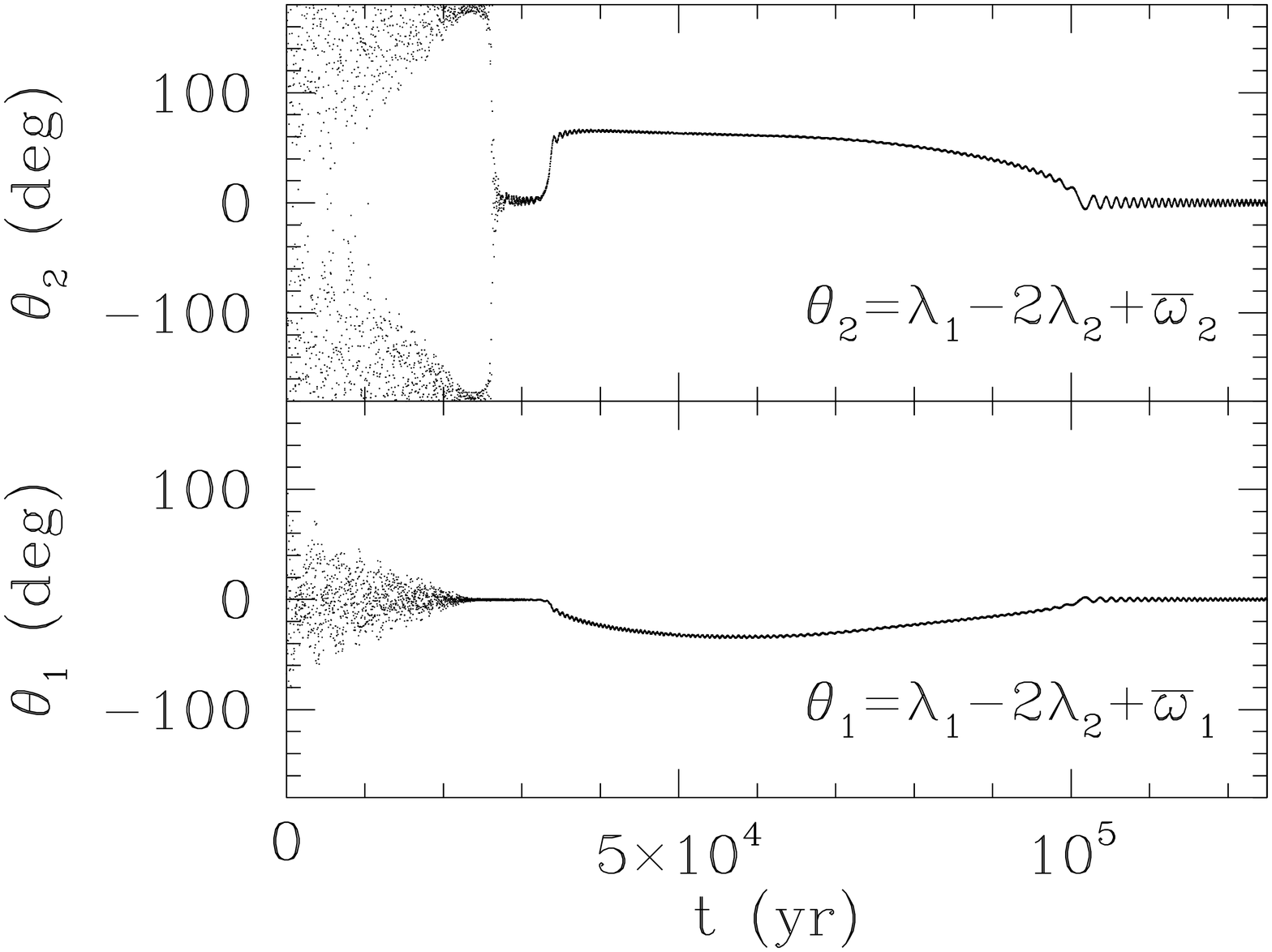}
\caption{Evolution into 2:1 resonances for a calculation with
$m_1/m_2 = 1.5$ and $(m_1 + m_2)/m_0 = 10^{-3}$.}
\end{figure}

\begin{figure}
\plottwo{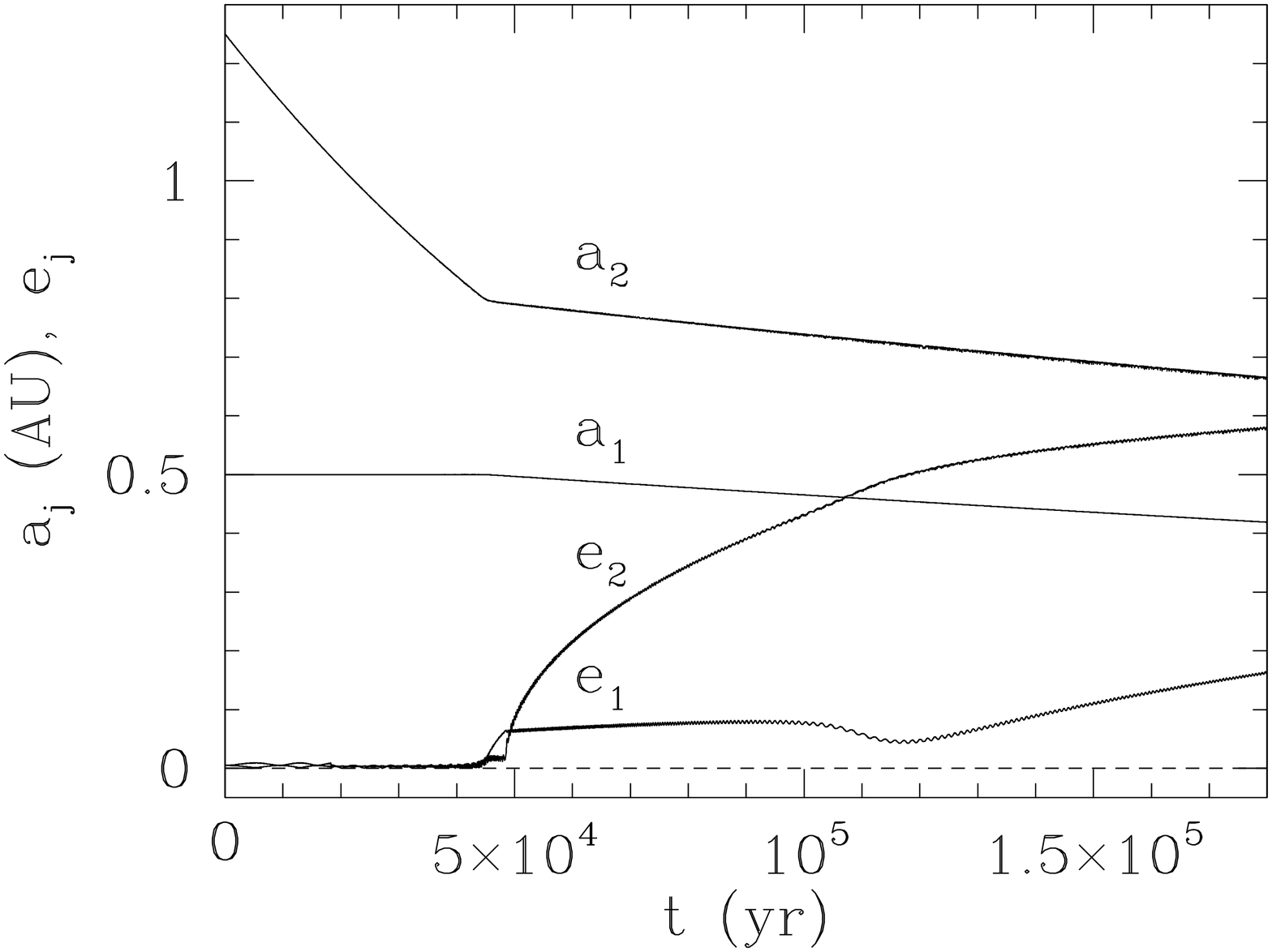}{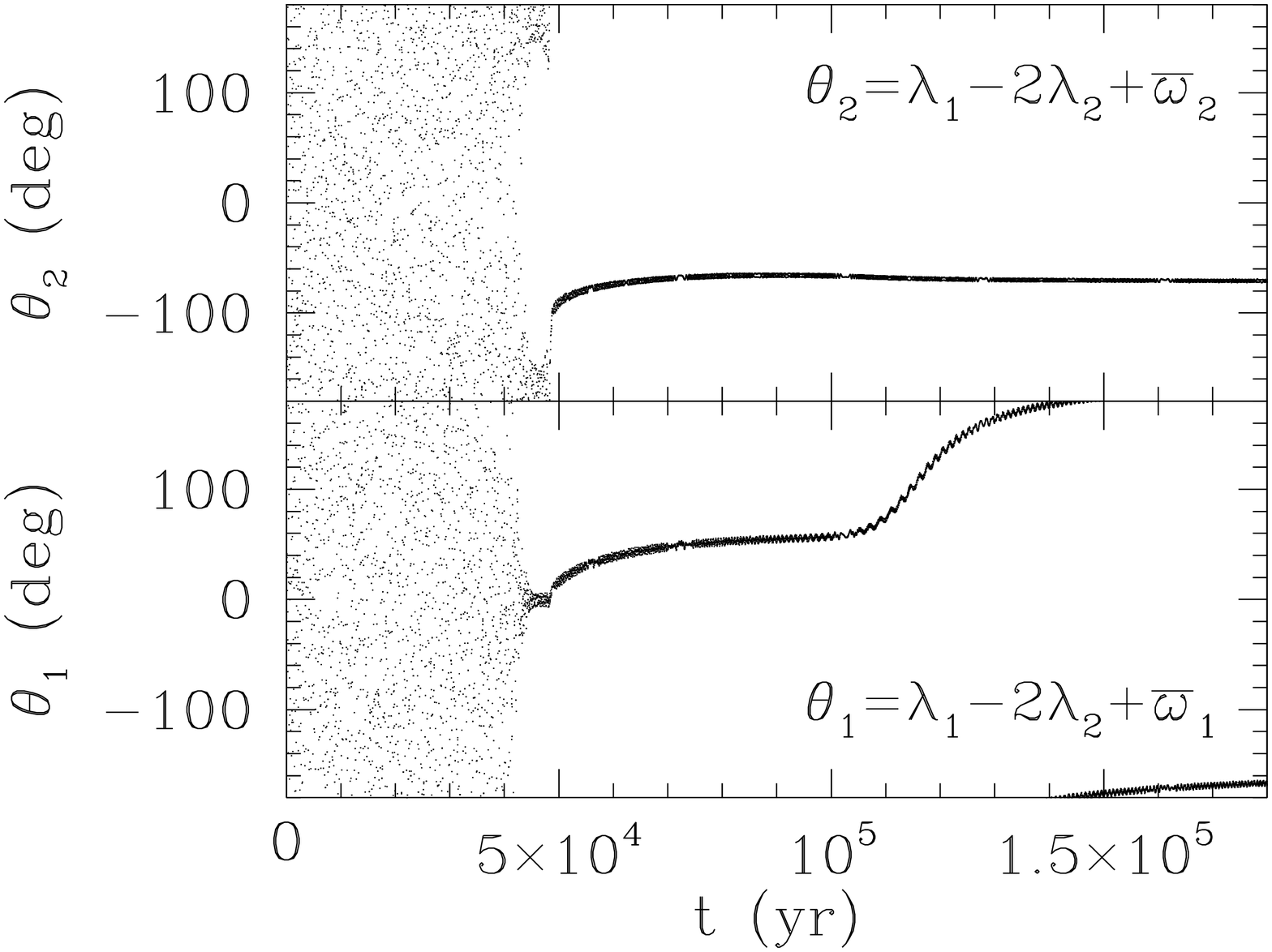}
\caption{Same as Fig.~2, but for a calculation with $m_1/m_2 = 4.16$.}
\end{figure}

The likely origin of the resonances in the GJ\,876 system is the
differential planet migration due to planet-nebula interaction.
The planets are sufficiently massive to open gaps in the nebula
disk surrounding the young GJ\,876 and to clear the disk material between
them (Bryden et al. 2000; Kley 2000), and the resulting planet-nebula
interaction typically forces the outer planet to migrate inward on the
disk viscous timescale (e.g., Ward 1997).
Figure 1 shows the evolution of the semimajor axes $a_j$, eccentricities
$e_j$, and resonance variables $\theta_1$ and $\theta_2$ for a calculation
where the outer planet of the GJ\,876 system is forced to migrate inward
and there is no eccentricity damping.
Both $\theta_1$ and $\theta_2$ are captured into libration about $0^\circ$,
but the $e_j$ quickly exceed the observed values (dotted lines) with
continued migration within the resonance.
Therefore, unless resonance capture occurred just before nebula dispersal,
eccentricity damping is required to produce the observed $e_j$ of the
GJ\,876 system.
Eccentricity damping of the form ${\dot e}_2/e_2 = -K|{\dot a}_2/a_2|$
causes the eccentricities to reach constant values that decrease with
increasing $K$.
Figure 1 also shows the evolution of the $e_j$ for a calculation with
$K = 100$, which is chosen to yield the observed $e_j$.
In Figure 1 we have neglected the outward migration of the inner planet
due to the torques induced by any nebula left on the inside of the inner
planet, as this does not qualitatively change the evolution except to
require less eccentricity damping ($K \approx 10$) to arrive at the
observed eccentricities (Lee \& Peale 2002).

\section{Diversity of 2:1 Resonance Configurations}

As the example in Figure 1 shows, differential migration of planets
through capture into a set of resonances and subsequent evolution without
eccentricity damping cause the system to pass through a continuous
sequence of stable libration states, whose configurations depend on the
values of the eccentricities $e_j$ and the mass ratios $m_1/m_2$ and
$(m_1+m_2)/m_0$ (where $m_0$ is the mass of the primary).
Where a system with a given set of parameters is left along this sequence
depends on the termination of the migration through nebula dispersal,
or much more probably, the magnitude of the eccentricity damping.

For $(m_1+m_2)/m_0 \approx 10^{-3}$ (a factor of 10 smaller than in the
GJ\,876 system), $m_1/m_2 \la 1$, and very small initial $e_j$,
a system is captured initially into a Io-Europa like configuration,
with $\theta_1$ librating about $0^\circ$ and $\theta_2$ librating about
$180^\circ$.
Continued migration forces $e_1$ to larger values and $e_2$ from
increasing to decreasing until the system passes smoothly over to
the GJ\,876 configuration with all $\theta_j$ librating about $0^\circ$.
Then both $e_1$ and $e_2$ increase smoothly as in Figure 1 until
instability ensues at very large $e_j$.
For $1 \la m_1/m_2 \la 2.5$ (see Fig.~2), the passage from the Io-Europa
configuration to the GJ\,876 configuration is as above, but when the
increasing $e_1$ reaches $\sim 0.1$--$0.3$ (depending on $m_1/m_2$),
the libration centers depart by tens of degrees from $0^\circ$, leading to
stable libration of the $\theta_j$ far from either $0^\circ$ or $180^\circ$.
This asymmetric libration can persist for a wide range of $e_1$, but the
system returns to the configuration with all $\theta_j$ librating about
$0^\circ$ for sufficiently large $e_1$.
If $m_1/m_2 \ga 2.5$ (see Fig.~3), the resonance variables never reach the
GJ\,876 configuration after leaving the Io-Europa configuration.
Within each type of evolution, the details of the path
depend only on $m_1/m_2$ and not on the total planetary mass, provided
$m_j$ are not too large.  
If $(m_1+m_2)/m_0 \approx 10^{-2}$ as in the GJ\,876 system, fairly large
$e_j$ are generated before resonance encounter and the Io-Europa
configuration expected at small $e_j$ is skipped (see Fig.~1).

\section{The 55\,Cnc System and 3:1 Resonance Configurations}

\begin{figure}
\plottwo{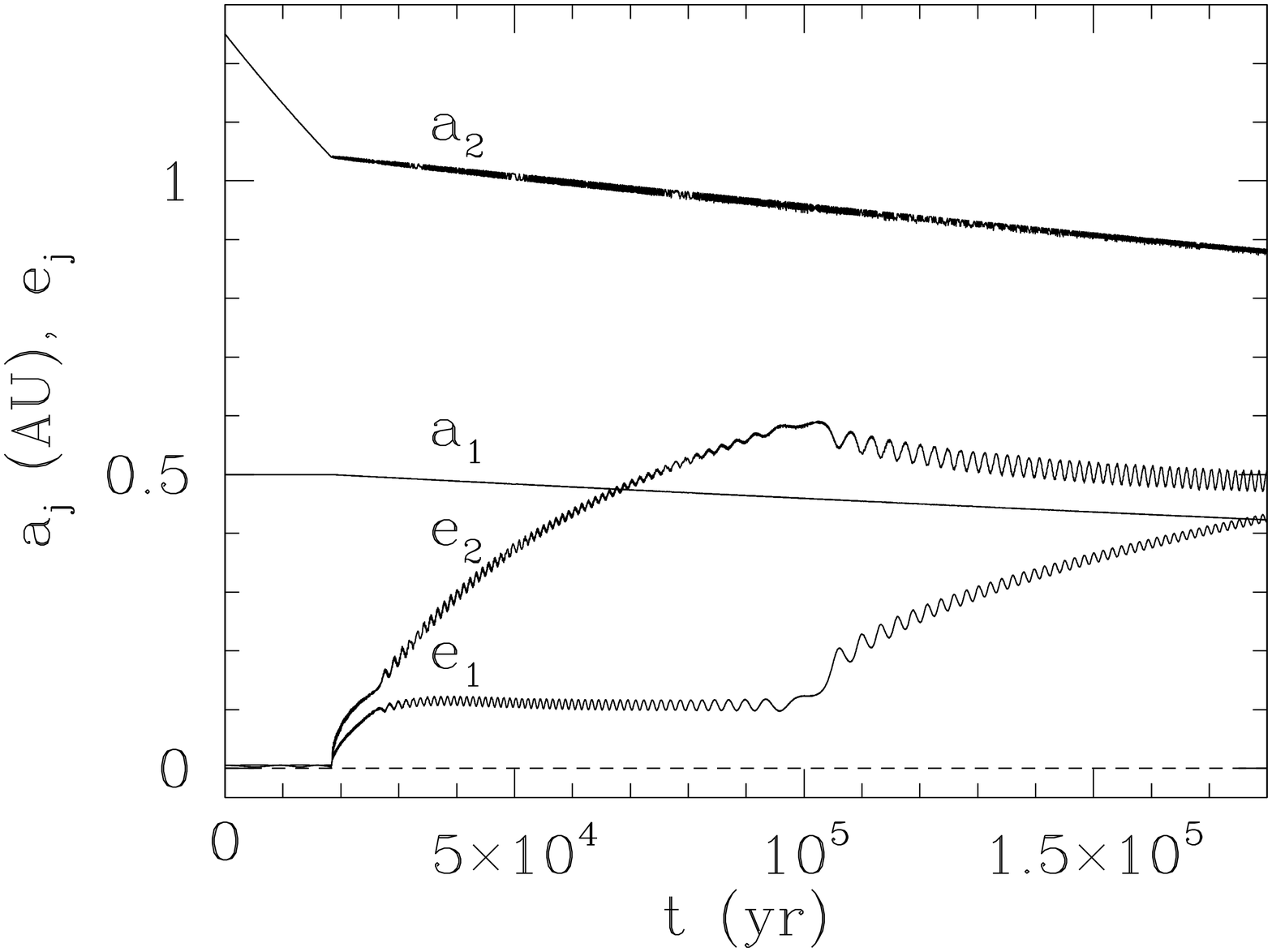}{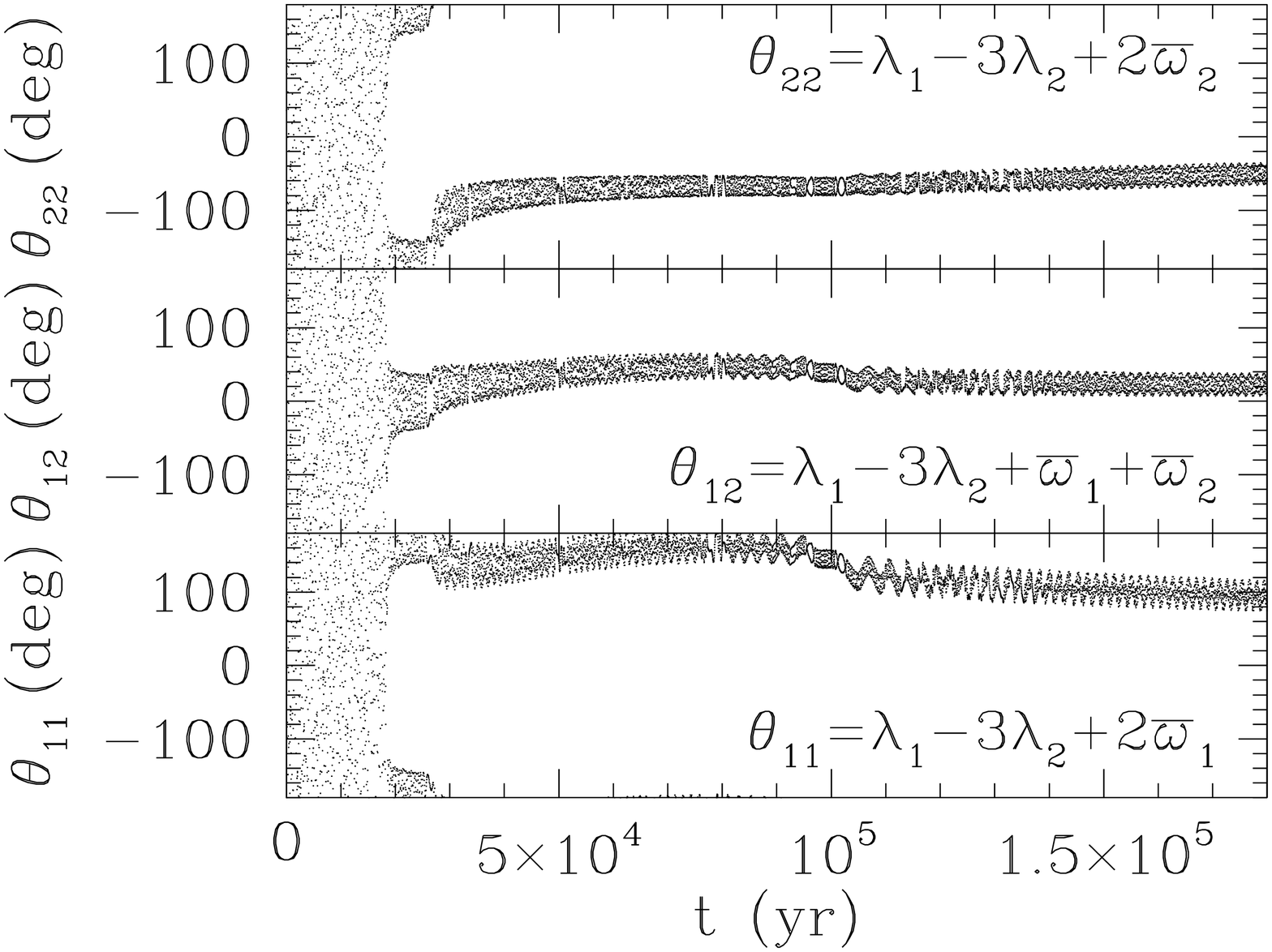}
\caption{Evolution into 3:1 resonances for a calculation with the same
masses as in Fig.~3.}
\end{figure}

For the inner two planets of the 55\,Cnc system [with minimum
$(m_1+m_2)/m_0 \approx 10^{-3}$] and the expected range of migration rates,
capture into the 3:1 resonances is sensitive to the initial conditions
and migration rate.
Figures 3 and 4 show the results of two calculations that are different
only in the initial orientation of the orbits, and the former is captured
into 2:1 resonances while the latter is captured into 3:1.
As shown in Figure 4, there are also 3:1 resonance configurations with
asymmetric librations.
The ``normal'' behavior for small eccentricities where the 3:1 resonance
variables (see Fig.~4 for definitions) librate about $0^\circ$ or
$180^\circ$ changes to librations about angles far removed from either of
these as the system is forced deeper into the resonances with continued
migration.

\acknowledgments
This research was supported in part by NASA PGG grants NAG5 3646 and NAG5
11666.

\end{document}